# Direct Kerr-frequency-comb atomic spectroscopy

*Liron Stern*[\*,1,2], *Jordan R. Stone*[1,2], *Songbai Kang*[1,2], *Daniel C. Cole*[1,2], *Myoung-Gyun Suh*[3], *Connor Fredrick*[1,2], *Zachary Newman*[1,2], *Kerry Vahala*[3], *John Kitching*[1], *Scott A Diddams*[1,2], and *Scott B. Papp*[\*,1,2]*

[1]Time and Frequency Division, National Institute for Standards and Technology, Boulder, CO 80305 USA
[2]Department of Physics, University of Colorado Boulder, Boulder, CO 80309 USA
[3]T. J. Watson Laboratory of Applied Physics, California Institute of Technology, Pasadena, California 91125 USA
*Corresponding authors: liron.stern@nist.gov and scott.papp@nist.gov*

**Microresonator-based soliton frequency combs– microcombs – have recently emerged to offer low-noise, photonic-chip sources for optical measurements. Owing to nonlinear-optical physics, microcombs can be built with various materials[1] and tuned or stabilized with a consistent framework[2]. Some applications require phase stabilization[3], including optical-frequency synthesis[4] and measurements[5], optical-frequency division[6], and optical clocks[7,8]. Partially stabilized microcombs can also benefit applications, such as oscillators[9], ranging[10], dual-comb spectroscopy[11,12], wavelength calibration[13,14], and optical communications[15]. Broad optical bandwidth, brightness, coherence, and frequency stability have made frequency-comb sources important for studying comb-matter interactions with atoms and molecules[16,17]. Here, we explore direct microcomb atomic spectroscopy, utilizing a cascaded, two-photon 1529-nm atomic transition of rubidium. Both the microcomb and the atomic vapor are implemented with planar fabrication techniques to support integration. By fine and simultaneous control of the repetition rate and carrier-envelope-offset frequency of the soliton microcomb, we obtain direct sub-Doppler and hyperfine spectroscopy of the $4^2D_{5/2}$ manifold. Moreover, the entire set of microcomb modes are stabilized to this atomic transition, yielding absolute optical-frequency fluctuations of the microcomb at the kilohertz-level over a few seconds and <1 MHz day-to-day accuracy. Our work demonstrates atomic spectroscopy with microcombs and provides a rubidium-stabilized microcomb laser source, operating across the 1550 nm band for sensing, dimensional metrology, and communication.**

Spectroscopy of atoms and molecules supports studies of quantum matter and numerous applications. An important advance in laser spectroscopy and metrology has been the use of optical-frequency combs. Specifically, combs can interact with atoms and molecules in Direct Frequency Comb Spectroscopy (DFCS), providing conceptually important measurements[16,17]. Indeed, in recent years DFCS has been applied to demonstrate sensitive and broadband molecular spectroscopy[18,19], breath analysis[20], precision atomic spectroscopy[21-23], steering Raman transitions[24], atomic clocks,[25] Ramsey spectroscopy[26], temperature sensing[27], and XUV spectroscopy[28], utilizing numerous techniques such as dual-comb spectroscopy, cavity-enhanced spectroscopy, fluorescence and Ramsey spectroscopy. Such a wide range of application directions and techniques are enabled by robust and controllable frequency-comb sources.

Recently, dissipative Kerr solitons in optical microresonators have been demonstrated to provide a compact platform for low-noise, microwave-rate, and low-power frequency combs[1]. Pumped with a continuous laser, stable soliton pulses can be generated in a Kerr-nonlinear microresonator. With these advantages, an unresolved question is how microcombs can be used in DFCS. Indeed, soliton microcombs have already been used for various applications[1] where for most it is essential to leverage precise frequency control of the soliton mode spectrum $v_m = f_{ceo} + m \cdot f_{rep}$ through the two degrees of freedom, i.e. the repetition frequency $f_{rep}$ and the carrier-envelope-offset frequency $f_{ceo}$. However, these parameters are coupled together in a complex manner by the soliton dynamics; this was determined in Ref. 2 for high-*Q* silica microresonators. Through the so-called 'fixed-points' of a microcomb, the possibility exists to decouple and precisely control $f_{rep}$ and $f_{ceo}$. We can use this feature for high resolution microcomb DFCS. A second approach is to stabilize $f_{rep}$, using phase modulation of the microcomb pump laser at the free-spectral range (FSR). This technique was explored in Ref. [29], and we can use it in DFCS to define a constant $f_{rep}$ with respect to a microwave clock, which completely decouples $f_{ceo}$ from $f_{rep}$ and allows tuning and stabilization.

A parallel benefit of microcomb DFCS is to derive a compact, low-power, and stable frequency-comb spectrum by stabilization to a DFCS signal. We can implement a chip-scale physics package by directly interfacing a microcomb and a rubidium alkali vapor cell. Indeed, in recent years the integration of alkali vapors and chip-scale photonic platforms has yielded interesting results, including anti-resonant reflecting optical waveguides (ARROW)[30], atomic cladding wave guides[31] and chip-integrated grating coupled cells[32].

Here we demonstrate microcomb DFCS of the cascaded 5P$_{3/2}$-4D$_{5/2}$ transition in [85]Rb at 1529.37 nm. This telecom-wavelength-band atomic transition is convenient and unique given the technological advantage of operating microcombs with telecom lasers and other components. We generate a soliton microcomb with a 1536-nm phase-modulated pump laser and use one microcomb mode to interact with Rb atoms in a micromachined vapor cell[33]. By also probing the near-infrared Rb D2 transition at 780 nm[34,35] with a second laser, we resolve the entire $4^2D_{5/2}$ hyperfine manifold composed of ~10 MHz linewidth atomic transitions. This double resonance optical pumping (DROP) technique[36] not only enables an enhancement in signal-to-noise ratio with DFCS, but is also insensitive to saturation from modes of the microcomb that are off-resonance of rubidium transitions. By frequency locking the microcomb $f_{ceo}$ to the F = 4 to F' = 3 hyperfine transition amongst the $4^2D_{5/2}$ manifold, we stabilize the fractional-frequency noise at the 10$^{-11}$ level. Moreover, we perform a day-to-day accuracy assessment, which indicates a <1-MHz repeatability in the absolute $f_{ceo}$ of the microcomb. The microcomb DFCS technique we describe is a general, robust, and compact approach to implement an atomic-transition-referenced microcomb, and it could be expanded to other spectral ranges and to multi-photon DFCS.

Figure 1 presents an overview of our apparatus and results. We generate a soliton microcomb with a silica resonator that is evanescently coupled via a tapered fiber[37]; see the schematic in Fig. 1a. The resonator is pumped with a 1536-nm laser, and the generated soliton microcomb spectrum extends approximately 60 nm across the telecom C-band. For microcomb DFCS, we send the soliton microcomb light into a Rb atomic vapor. We implement the microcomb-atom interaction via a pump-probe DROP scheme, involving one mode of the microcomb and a 780 nm laser; the atomic level diagram is depicted in the inset of Fig. 1a. The $5^2S$ ground state F = 3 hyperfine-state is primarily coupled to the $5^2P_{3/2}$ F = 3 cycling transition. This is achieved by using the D2 transition at 780 nm, which acts as a probe. A second transition at the telecom wavelength of 1529 nm, couples the $5^2P_{3/2}$ state to the $4^2D_{5/2}$ hyperfine manifold and acts as an atomic-pump (distinct from the frequency-comb pump). This pump-probe arrangement allows us to imprint the spectroscopic information of the telecom hyperfine manifold on the 780 nm probe. Moreover by nature of the optical-pumping process (i.e. transfer of atomic population between the two $5^2P_{3/2}$ ground states), we gain a substantial increase in signal due to decrease in population in the upper $5^2P_{3/2}$ state[36]. Both these features make the cascaded atomic system appealing for implementation of DFCS[38]. Figure 1b shows the devices that we use, namely a high-Q silica wedge resonator on a silicon chip and a micro-machined, dispenser-based Rb atomic vapor cell.

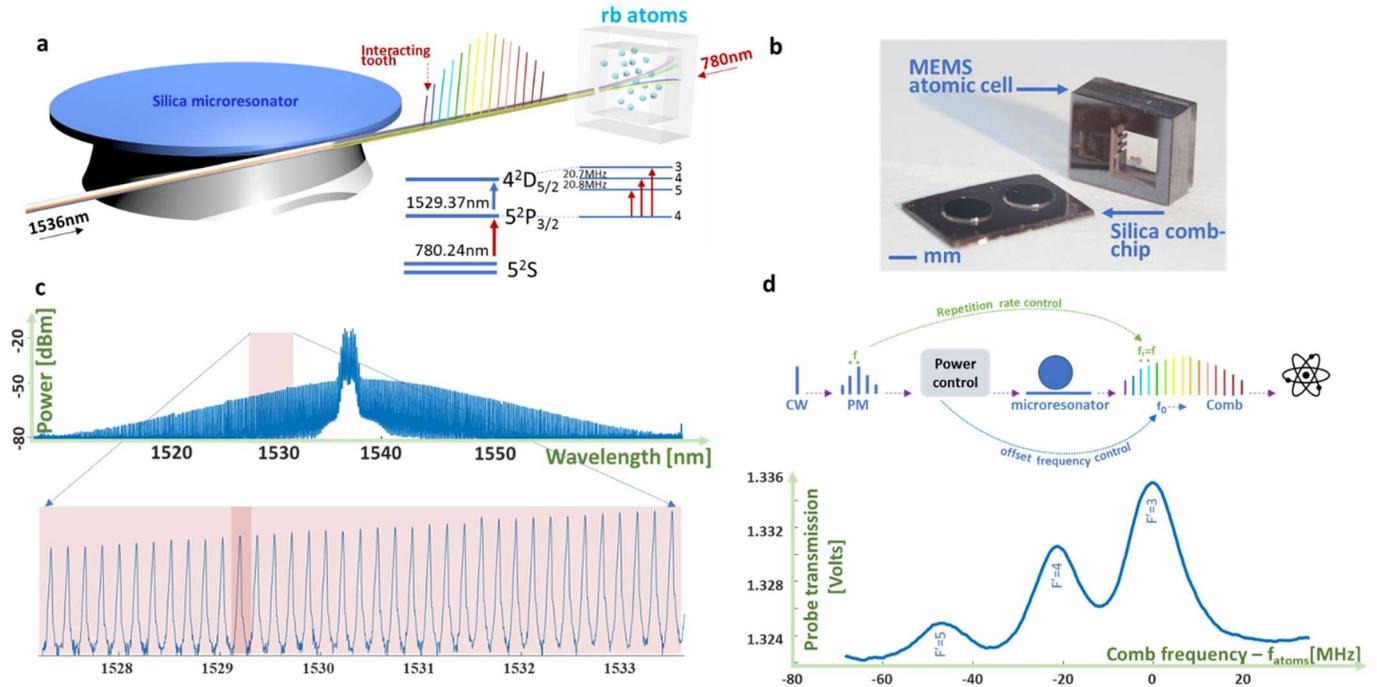

**Fig. 1: Direct Kerr-comb atomic spectroscopy.** a) Conceptual depiction of the microcomb and atomic systems. A 1536-nm pump laser energizes a single-soliton pulse in a silica whispering-galley-mode resonator, which in turn illuminates a millimeter-scale micro-machined rubidium cell. The comb has a 1529-nm mode that is resonant with a [85]Rb two-step atomic transition; see relevant Rb atomic-level scheme in inset. b) Photograph of typical Kerr-comb chip, and micro-machined Rb vapor cell. c) Soliton microcomb optical spectrum acquired with 20 pm resolution; zoomed section highlights the mode that interacts with the atomic medium. d) Top: Arrangement and physical mechanisms to initiate the soliton microcomb and control its $f_{\rm rep}$ and $f_{\rm ceo}$. Bottom: Probe transmission signal as function of comb frequency, revealing the $4^2D_{5/2}$ hyperfine manifold.

We create the soliton microcomb by pumping the silica resonator with a 1536-nm external-cavity diode laser (ECDL). Direct, deterministic single soliton pulses are generated by pumping the microresonator with 100 mW of phase modulated (PM) light[29]; the PM frequency is derived from a hydrogen-maser-referenced microwave synthesizer that is set close to the resonator's FSR of ~22 GHz. With the PM engaged, the microcomb's $f_{\rm rep}$ is tightly locked to the PM frequency. This PM technique allows us to initiate single solitons without transitioning through the chaotic regime. The optical-frequency spectrum is presented in Fig. 1c, where a comb spanning across the C-band is evident, as well as the seeding PM light. A zoomed section shows the specific comb mode that interacts with the atomic medium.

We use microcomb $f_{\rm ceo}$ tuning to acquire the DFCS signal of the cascaded Rb transitions. Operationally, we change the pump-laser power in order to exclusively tune $f_{\rm ceo}$, but this requires a detailed two-part microcomb control procedure. First, we rely on the PM pumping technique to maintain a fixed value of $f_{\rm rep}$. Second, we stabilize the frequency detuning of the pump laser with respect to the silica resonance, using an offset Pound-Drever-Hall (PDH) technique[2]. Changing the pump-laser power causes thermal and nonlinear shifts of the silica resonance and induces an exclusive shift in $f_{\rm ceo}$ of the entire equidistant comb, which we use for Rb DFCS. An illustration of this process is shown in Fig. 1d, along with spectroscopy data of the Rb transitions obtained with an approximately 50% change in the microcomb pump power. Specifically, the 1529-nm comb mode induces a variation in the 780 nm probe laser transmission through the Rb vapor cell. Three clear peaks (as anticipated from the DROP nature of this process) are evident corresponding to the three dipole-allowed transitions from the $5P_{3/2}$ F = 4 state to the $4D_{5/2}$ 3,4,5 states. The linewidth is measured to be ~10 MHz, which is close to the limit of 6 MHz imposed by the intermediate $5P_{3/2}$ natural lifetime.

Clearly, the ability to decouple the microcomb's physical controls, ie. pump-laser power and frequency, and the resonator detuning, from the microcomb's degrees of freedom, ie. $f_{\rm ceo}$ and $f_{\rm rep}$, is central to the implementation of high-resolution DFCS and subsequent stabilization. The coupling of these parameters arises in a complex manner from frequency shifts of the thermo-optic effect, the Kerr effect, and the self-soliton effect that all act simultaneously. Therefore, a key question is to what extent our microcomb operation procedure, which is shown with all components in Fig. 2a, achieves such decoupling. To answer this, we carry out an experiment in which we scan $f_{\rm ceo}$ and monitor $f_{\rm rep}$ with a microwave-frequency-counting system; see Fig. 2a. For microwave counting, we select only a portion of the comb separate from the PM pump modes. We present such a measurement in Fig. 2b for two different scenarios; the black- and blue-colored traces indicate $f_{\rm rep}$ and the orange trace is the pump power. In the first scenario, the pump power is kept constant and the $f_{rep}$ data indicates a stable lock. In the second scenario, the power is varied periodically with an amplitude corresponding to a 50 % change, but $f_{rep}$ remains unchanged indicating the robustness of the locking. To compare the scenarios in detail, we record a more extensive frequency-counter dataset and calculate the overlapping Allan deviation; see Fig. 2c. Indeed, when we abruptly vary the pump power, we do not observe a change in the locked accuracy or precision of the soliton microcomb repetition frequency. These data confirm the capability to decouple $f_{ceo}$ and $f_{rep}$, solving an important challenge for high precision soliton microcomb DFCS. By recognizing that such power variation induces the spectroscopy presented in Fig. 1d, we conclude that the soliton pulse envelope remains intact, whilst independently controlling of underlying carrier wave.

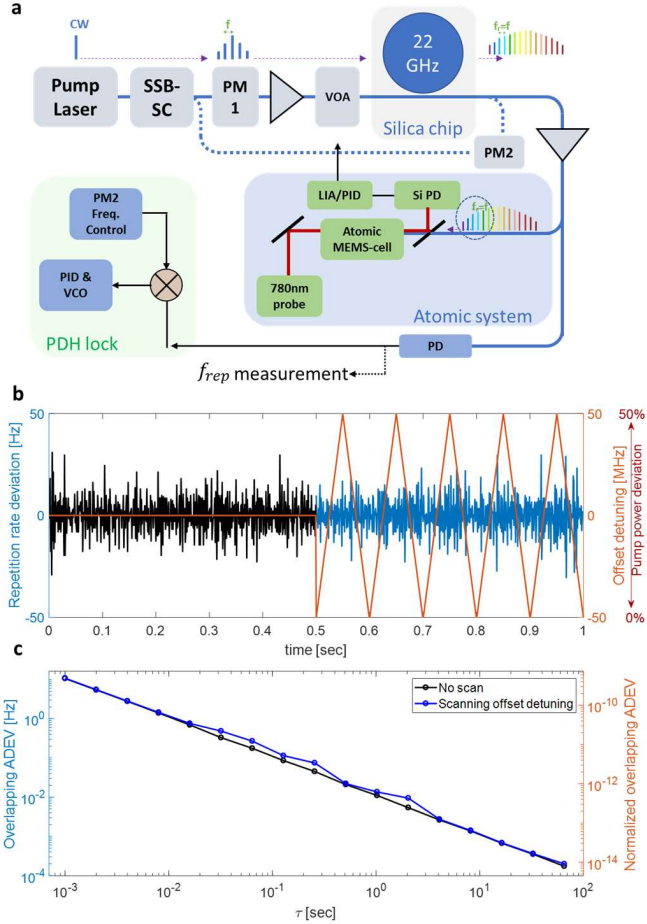

**Fig. 2: DFCS apparatus and demonstration of $f_{rep}$ control.** a) An ECDL drives a single-sideband, suppressed-carrier (SSB-SC) frequency shifter, driven by a voltage-controlled oscillator (VCO). The pump beam is phase modulated (PM1) at FSR. We implement a PDH servo-loop to control the resonator detuning with a counter propagating phased modulated (PM2) beam. A portion of the comb spectrum is amplified and sent to a micromachined atomic cell, which is also illuminated by a counter propagating CW 780 nm probe laser. By use of a dichroic mirror and a Si photodetector, the 780 nm light is monitored. Lock-in amplification and a PID servo locks $f_{ceo}$ to the atomic transition. b) Soliton microcomb $f_{rep}$ and pump power setting versus time (red). Blue and black $f_{rep}$ data indicates the different power conditions. c) Corresponding overlapping Allan deviation (ADEV) for the case of constant and swept pump power.

We demonstrate one specific use of microcomb DFCS: Stabilization of the absolute optical frequency of the entire set of comb teeth. In this experiment, we maintain the repetition-frequency stability, using PM pumping as described above. To implement a servo lock of $f_{ceo}$ with respect to the DFCS signal, we dither the pump-laser power to obtain an error signal and use a PID controller to provide feedback. The variable optical attenuator indicated in Fig. 2a provides the dither. Stabilization of the microcomb via the DROP spectroscopy technique also requires that the 780 nm probe laser is stabilized to the Rb D2 transition. In this work, we accomplish the 780-nm laser stabilization in a separate saturated absorption apparatus, but this could in principle also be performed in the same micromachined cell as the DROP spectroscopy.

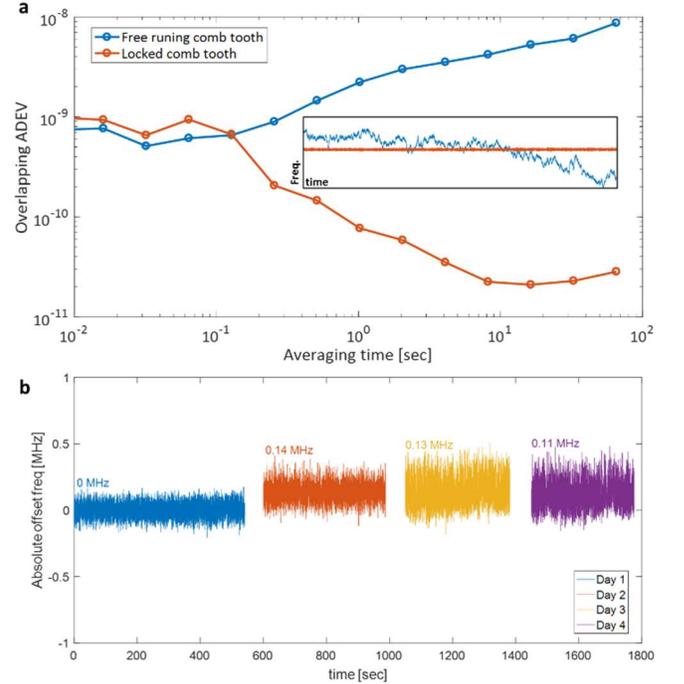

**Fig. 3: Direct Kerr-comb atomic referencing** a) Uncertainty assessment of the optical frequency precision, represented by an overlapping Allan deviation plot for the locked (red) and unlocked case (blue). The inset shows the time domain data over ~10 minutes from which the Allan deviation is calculated. b) Day to day optical frequency traces obtained by re-initiating the Kerr-comb and locking the comb directly to the cascaded atomic transition.

We characterize the frequency stability of the Rb-referenced microcomb by frequency counting a specific tooth (a few modes apart in frequency from the pump frequency) with respect to a self-referenced erbium-fiber frequency comb[39]. In Fig. 3a, we present the overlapping Allan deviation of the microcomb with Rb stabilization (red points) and without (blue points). By referencing the microcomb to the hyperfine structure of Rb, we improve its fractional-frequency stability to as low as $2 \times 10^{-11}$ after 10 seconds of measurement time. Therefore, with our system, all the microcomb mode frequencies are stable at the ~10 kHz level.

A potentially more important benefit of atomic-Rb stabilization of our 1550-nm-band soliton microcomb is long-term repeatability of the microcomb mode frequencies. In this manner, a microcomb could be available for an application without need for re-calibration or alignment. To assess day-to-day repeatability, we operate the system over a span of four days, each day restarting the microcomb and Rb laser system components, initiate a soliton microcomb, tune the microcomb into resonance of the Rb transition, apply the stabilization, and count the optical frequency. We leave the Rb temperature stabilization running at all times. We present such results in Fig. 3b, where each optical-frequency trace (indicated by different colors) represents a frequency measurement performed on a separate day. The maximum day-to-day frequency change is 140(4) kHz (from the first day to second), and the standard deviation of the shifts is 65(4) kHz. The absolute deviation corresponds to fractional uncertainty of $7 \cdot 10^{-10}$ and the last three days are a bit better. Considering that the system is subject to environmental perturbations, is unshielded from magnetic fields, and is not packaged, such stability and accuracy it would likely be possible to improve upon this performance. And yet, for the given system such accuracies and stability already support applications such as chip-scale based dimensional spectroscopy, hazardous material sensing, and wavelength meter calibration. Moreover, when considering a miniaturized system, such demonstrated level of performance for stabilization of $f_{rep}$ is consistent with commercial Rb microwave references and crystal oscillators.

To summarize, we have demonstrated direct Kerr-microcomb atomic spectroscopy by exploiting the cascaded $5P_{3/2}$-$4D_{5/2}$ transition in $^{85}$Rb. We use a comb mode near the $5P_{3/2}$-$4D_{5/2}$ transition frequency to interact with the $^{85}$Rb atoms by sending most of the comb spectrum directly into the micromachined vapor cell. Our spectroscopy experiments are enabled by fine frequency control of the microcomb, which allows us to resolve the closely spaced hyperfine manifold with linewidths of ~10 MHz. Even higher resolution spectroscopy is possible due to continuous sweeping of the microcomb. By implementing a servo-loop, we lock the comb to the atomic transition with a frequency uncertainty approaching $10^{-11}$. Accuracy assessment of the system shows a sub-megahertz accuracy over a few days. This demonstration paves the way for other exciting Kerr-based DFCS experiments.

## Methods:

### PM soliton initiation

The ECDL drives a single-sideband (SSB), suppressed-carrier frequency modulator controlled by a high-bandwidth voltage-controlled oscillator (VCO) for control of the pump laser. A portion of the laser is frequency shifted and modulated by means of an acousto-optic modulator and phase modulated, whereas one sideband of the phase-modulated light is locked to the cavity resonance. The high-bandwidth feedback allowed by the VCO/SSB scheme allows thermal instabilities associated with the red detuning that is required for soliton generation to be overcome. A counterpropagating (with respect to the AOM shifted beam) phase modulated pump beam is coupled to the resonator, thus parametrically seeding the resonator and allowing deterministic creation of single solitons. To generate single solitons, the pump laser is initially detuned ~25 linewidths from resonance, and subsequently the detuning is decreased to ~3 linewidths, where a soliton step is observed.

### Micromachined vapor cell fabrication

The vapor cell has been constructed by using a combination of wafer-scale fabrication silicon frames, anodic bonding and Rb dispenser pills[33]. First, a 3 mm thick silicon wafer is structured using a deep reactive ion etching process, to create a silicon frame with two chambers connected by channels. The large chamber has the dimensions of 3 mm X 3 mm, whereas the small chamber has the dimension of 1.5 mm X 1.5 mm. Next, a 0.7 mm thick Pyrex window is anodically bonded to the silicon frame, to form a preform. Following, in a vacuum environment a Rb dispenser pill is introduced in the small enclosure, and a second Pyrex window is anodically bonded to the exposed side of the silicon frame, creating a closed cell. Finally, the cell is activated by illuminating the dispenser pill with approximately 1 W of 980 nm laser power, to release natural rubidium into the chambers enclosed within the cell.

### Spectroscopy & stabilization setup

The rubidium spectroscopy setup is based on the two-photon cascaded transition in $^{85}$Rb. A 780nm ECDL is locked to the F = 2 to F' = 3 D2 transition in a centimeter-scale rubidium cell using saturation spectroscopy. The frequency uncertainty of this laser characterized by comparing its frequency to a commercial frequency comb is $3 \cdot 10^{-12}/\tau^{1/2}$. The laser is sent over fiber to illuminate the mm-scale MEMS cell with a beam diameter of approximately 1 mm. This laser serves as a probe for the counter-propagating Kerr-comb light interacting with the atoms. The Kerr-comb has an approximate beam diameter of 900 μm, and the interacting comb mode has an amplified power of approximately 50 μW. The Kerr-comb is stabilized to the cascaded transition by feeding the VOA with approximately a 500 Hz demodulated error signal. Finally, a mode close to the pump frequency is filtered out, compared to a tooth of the commercial locked frequency comb system, and frequency counted. Through a separate systematic characterization of the DROP atomic transition, using a telecom ECDL, we assess two dominate contributions to frequency shifts in our system: variations in the cell temperature and optical power of the atomic system's pump and probe. This analysis shows that our cell has a normalized frequency temperature coefficient of $10^{-10}$/°C, and a frequency power coefficient of $\sim 2 \cdot 10^{-11}$ and $\sim 8 \cdot 10^{-11}$ for a change of 1 % of the 780 nm and 1529 nm powers, respectively.

## References:

1. Kippenberg, T. J., Gaeta, A. L., Lipson, M. & Gorodetsky, M. L. Dissipative Kerr solitons in optical microresonators. *Science* **361**, eaan8083 (2018).
2. Stone, J. R. *et al.* Thermal and Nonlinear Dissipative-Soliton Dynamics in Kerr-Microresonator Frequency Combs. *Phys. Rev. Lett.* **121**, (2018).
3. Briles, T. C. *et al.* Interlocking Kerr-microresonator frequency combs for microwave to optical synthesis. *Opt. Lett.* **43**, 2933 (2018).
4. Spencer, D. T. *et al.* An optical-frequency synthesizer using integrated photonics. *Nature* **557**, 81–85 (2018).
5. Lamb, E. S. *et al.* Optical-Frequency Measurements with a Kerr Microcomb and Photonic-Chip Supercontinuum. *Phys. Rev. Appl.* **9**, 024030 (2018).
6. Drake, T. E. *et al.* A Kerr-microresonator optical clockwork. *arXiv* (2018).
7. Papp, S. B. *et al.* Microresonator frequency comb optical clock. *Optica* **1**, 10 (2014).
8. Newman, Z. L. *et al.* Photonic integration of an optical atomic clock. *arXiv* (2018).
9. Liang, W. *et al.* Stabilized C-Band Kerr Frequency Comb. *IEEE Photonics J.* **9**, 1–11 (2017).
10. Suh, M.-G. & Vahala, K. J. Soliton microcomb range measurement. *Science* **359**, 884–887 (2018).
11. Dutt, A. *et al.* On-chip dual-comb source for spectroscopy. *Sci. Adv.* **4**, e1701858 (2018).
12. Suh, M.-G., Yang, Q.-F., Yang, K. Y., Yi, X. & Vahala, K. J.

12. Microresonator soliton dual-comb spectroscopy. *Science* **354**, 600–603 (2016).
13. Obrzud, E. *et al.* A microphotonic astrocomb. *Nat. Photonics* **13**, 31–35 (2019).
14. Suh, M.-G. *et al.* Searching for exoplanets using a microresonator astrocomb. *Nat. Photonics* **13**, 25–30 (2019).
15. Pfeifle, J. *et al.* Coherent terabit communications with microresonator Kerr frequency combs. *Nat. Photonics* **8**, 375–380 (2014).
16. Stowe, M. C. *et al.* Direct frequency comb spectroscopy. *Advances in Atomic, Molecular and Optical Physics* **55**, 1–60 (2008).
17. Marian, A., Stowe, M. C., Lawall, J. R., Felinto, D. & Ye, J. United time-frequency spectroscopy for dynamics and global structure. *Science (80-. ).* (2004). doi:10.1126/science.1105660
18. Bernhardt, B. *et al.* Cavity-enhanced dual-comb spectroscopy. *Nat. Photonics* **4**, 55–57 (2010).
19. Yu, M. *et al.* Gas-Phase Microresonator-Based Comb Spectroscopy without an External Pump Laser. *ACS Photonics* **5**, 2780–2785 (2018).
20. Thorpe, M. J., Balslev-Clausen, D., Kirchner, M. S. & Ye, J. Cavity-enhanced optical frequency comb spectroscopy: application to human breath analysis. *Opt. Express* **16**, 2387 (2008).
21. Barmes, I., Witte, S. & Eikema, K. S. E. High-Precision Spectroscopy with Counterpropagating Femtosecond Pulses. *Phys. Rev. Lett.* **111**, 023007 (2013).
22. Heinecke, D. C. *et al.* Optical frequency stabilization of a 10 GHz Ti:sapphire frequency comb by saturated absorption spectroscopy in $^{87}$rubidium. *Phys. Rev. A* **80**, 053806 (2009).
23. Stowe, M. C., Cruz, F. C., Marian, A. & Ye, J. High Resolution Atomic Coherent Control via Spectral Phase Manipulation of an Optical Frequency Comb. *Phys. Rev. Lett.* **96**, 153001 (2006).
24. Solaro, C., Meyer, S., Fisher, K., DePalatis, M. V. & Drewsen, M. Direct Frequency-Comb-Driven Raman Transitions in the Terahertz Range. *Phys. Rev. Lett.* **120**, 253601 (2018).
25. Gerginov, V., Tanner, C. E., Diddams, S. A., Bartels, A. & Hollberg, L. High-resolution spectroscopy with a femtosecond laser frequency comb. *Opt. Lett.* **30**, 1734 (2005).
26. Morgenweg, J., Barmes, I. & Eikema, K. S. E. Ramsey-comb spectroscopy with intense ultrashort laser pulses. *Nat. Phys.* **10**, 30–33 (2014).
27. Klose, A., Ycas, G., Cruz, F. C., Maser, D. L. & Diddams, S. A. Rapid, broadband spectroscopic temperature measurement of $CO_2$ using VIPA spectroscopy. *Appl. Phys. B* **122**, 78 (2016).
28. Cingöz, A. *et al.* Direct frequency comb spectroscopy in the extreme ultraviolet. *Nature* **482**, 68–71 (2012).
29. Cole, D. C. *et al.* Kerr-microresonator solitons from a chirped background. *Optica* **5**, 1304 (2018).
30. Yang, W. *et al.* Atomic spectroscopy on a chip. *Nat. Photonics* **1**, 331–335 (2007).
31. Stern, L., Desiatov, B., Goykhman, I. & Levy, U. Nanoscale light–matter interactions in atomic cladding waveguides. *Nat. Commun.* **4**, 1548 (2013).
32. Hummon, M. T. *et al.* Photonic chip for laser stabilization to an atomic vapor with $10^{-11}$ instability. *Optica* **5**, 443 (2018).
33. Liew, L.-A. *et al.* Microfabricated alkali atom vapor cells. *Appl. Phys. Lett.* **84**, 2694–2696 (2004).
34. Moon, H. S., Lee, W.-K. & Suh, H. S. Hyperfine-structure-constant determination and absolute-frequency measurement of the Rb $4D_{3/2}$ state. *Phys. Rev. A* **79**, 062503 (2009).
35. Moon, H. S., Lee, L. & Kim, J. B. Double-resonance optical pumping of Rb atoms. *J. Opt. Soc. Am. B* **24**, 2157 (2007).
36. Moon, H. S., Lee, L. & Kim, J. B. Double-resonance optical pumping of Rb atoms. *J. Opt. Soc. Am. B* **24**, 2157 (2007).
37. Yi, X., Yang, Q.-F., Yang, K. Y., Suh, M.-G. & Vahala, K. Soliton frequency comb at microwave rates in a high-Q silica microresonator. *Optica* **2**, 1078 (2015).
38. Moon, H. S., Ryu, H. Y., Lee, S. H. & Suh, H. S. Precision spectroscopy of Rb atoms using single comb-line selected from fiber optical frequency comb. *Opt. Express* **19**, 15855 (2011).
39. Ycas, G., Osterman, S. & Diddams, S. A. Generation of a 660–2100 nm laser frequency comb based on an erbium fiber laser. *Opt. Lett.* **37**, 2199 (2012).

**Acknowledgments:**

We thank Su-Peng Yu and Matthew Hummon for comments on the manuscript. The authors would like to acknowledge the Kavli Nanoscience Institute. This work is a contribution of the US government and is not subject to copyright in the United States of America.